# Recent advances of transition radiation: fundamentals and applications

Ruoxi Chen[1,2], Zheng Gong[1,2], Jialin Chen[1,2,3], Xinyan Zhang[1,2], Xingjian Zhu[4], Hongsheng Chen[1,2,5,6,*], and Xiao Lin[1,2,*]

[1]*Interdisciplinary Center for Quantum Information, State Key Laboratory of Extreme Photonics and Instrumentation, ZJU-Hangzhou Global Scientific and Technological Innovation Center, College of Information Science & Electronic Engineering, Zhejiang University, Hangzhou 310027, China.*

[2]*International Joint Innovation Center, the Electromagnetics Academy at Zhejiang University, Zhejiang University, Haining 314400, China.*

[3]*Department of Electrical and Computer Engineering, Technion-Israel Institute of Technology, Haifa 32000, Israel.*

[4]*School of Physics, Zhejiang University, Hangzhou 310027, China.*

[5]*Key Laboratory of Advanced Micro/Nano Electronic Devices & Smart Systems of Zhejiang, Jinhua Institute of Zhejiang University, Zhejiang University, Jinhua 321099, China.*

[6]*Shaoxing Institute of Zhejiang University, Zhejiang University, Shaoxing 312000, China.*

*Corresponding authors: xiaolinzju@zju.edu.cn (X. Lin); hansomchen@zju.edu.cn (H. Chen)

**Transition radiation is a fundamental process of light emission and occurs whenever a charged particle moves across an inhomogeneous region. One feature of transition radiation is that it can create light emission at arbitrary frequency under any particle velocity. Therefore, transition radiation is of significant importance to both fundamental science and practical applications. In this paper, we provide a brief historical review of transition radiation and its recent development. Moreover, we pay special attention to four typical applications of transition radiation, namely the detection of high-energy particles, coherent radiation sources, beam diagnosis, and excitation of surface waves. Finally, we give an outlook for the research tendency of transition radiation, especially its flexible manipulation by exploiting artificially-engineered materials and nanostructures, such as gain materials, metamaterials, spatial-temporal materials, meta-boundaries, and layered structures with a periodic or non-periodic stacking.**

Free-electron radiation originates from the particle-matter interaction and is a fundamental process of light emission [1-8]. Since free-electron radiation is able to create light emission at arbitrary frequency, it is of paramount importance to numerous applications [9-15], including high-energy particles detector, particle-beam diagnosis, free-electron lasers, high-power microwave/terahertz sources, electron microscopy, biomedical imaging, medical therapy, optical communications, and security.

Due to the exotic particle-matter interactions, the free-electron radiation could occur under different scenarios. Correspondingly, there are various types of free-electron radiation, including Cherenkov radiation [16-19], transition radiation [20-23], Smith-Purcell radiation [24-28], bremsstrahlung radiation [29-34], and synchrotron radiation [35-40], as shown in Fig. 1. Cherenkov radiation is the most famous type of free-electron radiation, and it occurs in a homogeneous matter only when a charged particle moves with a velocity larger than the Cherenkov threshold, namely the phase velocity of light in that matter [41-45]. Different from Cherenkov radiation, the occurrence of transition radiation has no specific requirement on the particle velocity, and it could happen whenever a charged particle moves across an inhomogeneous region [46-50], such as an optical interface. Smith-Purcell radiation emerges when a charged particle moves parallel and close to the surface of an optical diffraction grating [24-28]. Bremsstrahlung radiation, also known as the braking radiation, would appear if the charge particle decelerates or accelerates [29-34]. Synchrotron radiation originates from the circular motion of charged particles [35-40]. Due to the complexity of particle-matter interactions, the free-electron radiation, such as transition radiation, is a subject of extensive researches over the last several decades and is still a hot topic [51-55].

Below, we focus on the discussion of transition radiation. A brief review of the development of transition radiation is provided, including its interesting history and recent advances. To highlight the importance of transition radiation in science and technology, several typical applications of transition radiation are discussed in depth, ranging from high-energy particle detectors [56-65], coherent radiation source [66-71], beam diagnostics [72-76], to excitation of surface plasmon [77-84]. Due to the recent rapid development in material science and nanotechnology, an outlook on how to flexibly control the behavior of transition radiation is given by exploiting exotic artificially-engineered materials and nanostructures [85-90], including gain materials, spatial-temporal materials, meta-boundaries, and layered structures with a random stacking.

**Brief development history of transition radiation**

We begin with a brief review of the development history of transition radiation, as summarized in Fig. 2. The simplest case of transition radiation was first proposed by Ginzburg and Frank in 1946 [20] by exploring the bombardment of an electron at the interface between vacuum and a metal. The theory of transition radiation was experimentally confirmed in the visible range by Goldsmith and Jelley in 1959 [91], after the analysis of the collected radiation fields, including their polarization, excitation function and absolute yield. In their setup, a Van de Graaff generator was adopted to produce a 5 mega-electron-volt (MeV) beam bunch, which was then injected towards the vacuum-metal interface. Soon after the discovery of transition radiation, the quantum effect of transition radiation was also investigated by Garibian in 1960 [92].

The development of transition radiation is closely related to the interesting history of Ferrell radiation [93], which is featured with a peak near the plasma frequency in the radiation spectrum. In 1958, Ferrell proposed an approximate theory for the radiation of plasma oscillation and claimed that under suitable circumstances, the plasma oscillations would give off electromagnetic radiation [93]. Since Ferrell radiation could facilitate the plasma-frequency measurement of metals [93], Ferrell radiation was soon experimentally observed in 1960 [94,95]. In 1961, Silin and Fetisov pointed out that Ferrell radiation is merely the transition radiation [96]. In 1962, Stern argued that Ferrell's method and Ginzburg and Frank's theory of transition radiation are two distinct ways to consider the same phenomenon [97]. Stern highlighted that the physical mechanism of Ferrell radiation was misinterpreted by Silin and Fetisov, and it is "a surface effect" [97], namely "the contribution of radiative surface plasma oscillation (SPO)" [98], instead of the bulk longitudinal plasma oscillation in their study. In 1969, Economou emphasized that "there are no radiative SPO in the present geometry" [98]. Actually, he preferred the explanation of transition radiation, after the mathematical analysis of the denominator in the expression of transition radiation near the peak of Ferrell radiation [98]. Due to the recent advances in plasmonics [99-105], it is now acknowledged that the radiative SPO is essentially a leaky mode, denoted as the Ferrell mode [84,97,98,106]. In 2022, the Ferrell radiation was re-investigated in the time domain and was found able to occur far beyond the formation time, since it is supported by a long tail of bulk plasmons following the electron's trajectory deep into the plasmonic medium [107]. This way, the plasmonic tail is capable to mix surface and bulk effects and provides a sustained channel for electron-interface interaction. This time-domain finding may settle the historical debate in Ferrell radiation, regarding whether it is a surface or bulk effect, from transition radiation or plasmonic oscillation. Moreover, with the aid of Ferrell modes in uniaxial epsilon-near-zero materials, such as a thin hexagonal boron nitride (*h*BN) slab, an exotic phenomenon of low-velocity-favored transition radiation was recently proposed [108]. When the low-velocity-favored transition radiation occurs, the light emission from ultralow-energy particles with extremely-low velocities could exhibit comparable intensity as that from high-energy particles [108].

Due to the recent advances in nanofabrication, artificially-engineered materials or nanostructures, such as metamaterials [109-119], 2D materials [120-128], and photonic crystals [129-135], begin to play an important role in the flexible manipulation of transition radiation.

In 2009, Ref. [136] analyzed the case that a charged particle crosses an interface between a positive-index medium and a negative-index medium. Under this scenario, the transition radiation from the interface and the reversed Cherenkov radiation from the negative-index material would interfere and make the light emission complex. In 2012, Ref. [137] further studied the transition radiation from an average zero-index metamaterial, comprised of periodically alternating negative-index and positive-index layers. A strong radiation enhancement up to three orders of magnitude was predicted, due to the gigantic increase in the density of states at the positive-index/negative-index interface.

In 2012, Ref. [138] used graphene to tailor the transition radiation. Although the graphene is atomically thin, the free electron can efficiently excite graphene plasmons with probabilities in the order of one per electron [139]. In addition to the monolayer graphene [78,84,140-142], Ref. [142] also investigated the photonic and plasmonic transition radiation from multilayer graphene.

Due to the resonance effect, the analysis of transition radiation from periodic structures [143-147] is rather complicated. The related research could be dated back to the transition radiation in a periodically stratified plasma [148]. In 2003, Ref. [149] revealed that the behavior of Cherenkov radiation inside a two-dimensional photonic crystal is intrinsically coupled with the transition radiation and would appear without the Cherenkov threshold. In 2018, Ref. [150] further revealed the connection between Cherenkov radiation and transition radiation. Actually, Ref. [150] proposed a new mechanism – by exploiting the resonance transition radiation from one-dimensional photonic crystals – to create the effective Cherenkov radiation. This mechanism is capable to control the effective Cherenkov angles with high sensitivity, under any desired range of particle momentum, and is thus promising for the design of novel particle detectors.

**High-energy particle detectors based on transition radiation**

Particle detectors provide a powerful tool to detect, track and identify high-energy particles [151-155]. In high-energy physics, the charged particle could have a Lorentz factor $\gamma = 1/\sqrt{1-v^2/c^2}$ up to $10^4$, where $v$ is the particle velocity and $c$ is the speed of light in free space. For such high-energy particles, the sensitivity of common particle detectors (e.g. Cherenkov detectors [156-159]) is generally very low, and thus their detection is full of challenges. To address this issue, Garibian expanded the theory of transition radiation into the X-ray regime in 1959 and found the radiation energy is linearly proportional to the Lorentz factor $\gamma$, along with the radiation peak emerging at $\theta_{\max} = \gamma^{-1}$, if $\gamma \gg 1$ [160,161]. Therefore, the unique relation between the radiation energy and the Lorentz factor offers a new route to detect ultra-relativistic particles, even when their kinetic energy is up to tera-eV (TeV). Correspondingly, the particle detector based on the X-ray transition radiation is now known as the transition radiation detector [56-65].

The transition radiation detector, along with other particle detectors, has made a significant contribution to many famous experiments and the finding of new particles (e.g. W and Z bosons [162], the Higgs boson [163]). Actually, the transition radiation detector is widely used in many particle-physics laboratories (e.g. CERN, Femi Lab), as exemplified in Fig. 3. For example, in the H1 experiment, electron and positron were identified at the HERA electron proton collider [56]. In the experiment E799, the Fermi lab designed a transition radiation detector with a large aperture to provide π/e rejection [57]. In the NOMAD experiment, transition radiation detector discriminated electron and pion and could search for the $v_\mu \rightarrow v_\tau$ oscillation [58]. In the ALICE experiment, the transition radiation detector is served as a trigger on high $p_t$ $e^+ e^-$ pairs to reduce the collision rate to the readout event rate [59]. In the super proton synchrotron (SPS) of CERN, an inorganic scintillator-based Compton-scatter transition radiation detector in Fig. 3a is designed to detect high-energy electrons with an ultra-high Lorentz factor [60]. In the ATLANS experiment, the transition radiation tracker is a fast detector with thin detector layers realized with straw tubes and could provide both tracking information and particle identification; see its cutaway view in Fig. 3b [61]. In the compressed baryonic matter (CBM) experiment, the transition radiation detector is used to measure common hadrons, such as low-mass dileptons, charmed hadrons, and multistrange baryons [62].

The performance of transition radiation detectors is mainly determined by the properties of the associated radiator and detector. For experiments with relatively-low particle multiplicity, transition radiation detectors generally use multi-wire proportional chambers (MWPC) or straw tubes, filled with a Xenon based gas mixture to efficiently absorb the emitted photons [63]. For experiments with relatively-high particle multiplicity, the efficiency of traditional transition radiation detectors would decrease significantly due to the channel occupancy. Ref. [64] showed that the performance of transition radiation detectors could be improved by replacing MWPC or straw tubes with a high-granularity-micro-pattern gas detector, such as gas electron multipliers as shown in Fig. 3c. In addition, Ref. [150] found that the one-dimensional photonic crystal could create the resonance transition radiation with ultrahigh directivity in Fig. 3d and acts as a new type of radiator.

Correspondingly, the photonic crystal offers a promising versatile platform well suited for the identification of particles at high energy with enhanced sensitivity.

**Coherent light sources based on transition radiation**

If the electron beam has a length much shorter than the working wavelength, all electrons can be considered to emit in phase. As a result, the related transition radiation is coherent. The phenomenon of coherent transition radiation was first observed in 1991 [164]. One feature of the coherent transition radiation is that its energy could be enhanced by a factor of $N$ than that of the incoherent transition radiation, where $N$ is the electron number.

For any type of free-electron radiation, it takes a finite space domain instead of a point for photons to be emitted. This finite space domain is now known as the formation zone [165], which is a useful concept for the coherent transition radiation. In the context of free-electron radiation, this concept was first presented by Ter-Mikaelian in Landau's seminar in 1952 [166], and later developed by Landau himself [167,168], with its experimental confirmation in the 1990s. Later, Ginzburg extended this concept into transition radiation [169]. The concept of formation zone provides valuable guidance for practical applications based on the transition radiation. Particularly, the influence of formation zone is generally avoided in the design of transition radiation detectors [170].

According to Ginzburg's work [169], the length of the formation zone for the transition radiation, namely the formation length $L_f$, is defined as the distance that the charge field $E^q$ and the radiation field $E^R$ separate from each other. In other words, the contribution of the interference term $E^q \cdot E^R$ to the total field energy (proportional to $|E^q + E^R|^2$) is very small outside the formation zone. According to the definition, we have

$$L_\mathrm{f} = \frac{2\pi}{\left|\omega/v \pm \omega/c \sqrt{1-\frac{k_\perp^2 c^2}{\omega^2}}\right|} \qquad (1)$$

where $\omega$ is the angular frequency, $k_\perp$ is the component of wavevector perpendicular to the electron velocity, and $\pm$ correspond to the forward and backward radiation, respectively. Ginzburg's estimation on the formation zone is actually only applicable to the photon emission during the process of transition radiation. In 2017, along the line of Ginzburg's thought, the concept of formation zone was extended to the emitted surface plasmons during the process of transition radiation [78].

With the knowledge of formation zone, the generation of coherent transition radiation mainly relies on two ways. One way is to use an electron beam to directly bombard the optical interface. For example, Ref. [66] shows that the coherent terahertz (THz) radiation could be generated by an electron bunch with a duration in the order of tens of femtoseconds to picoseconds penetrating through the plasma-vacuum interface; see Fig. 4a. Such electron bunches are produced by a laser-plasma accelerator [66]. Similarly, the THz coherent transition radiation with energies in the order of sub-mJ/pulse could also be induced by letting the laser-driven electron beams cross a dielectric-vacuum interface [67] in Fig. 4b. When an electron beam has a femtosecond duration and hundreds of kiloampere peak current and penetrates through the plasma-vacuum interface, the terawatt ultraviolet coherent transition radiation could happen [68]; see the associated ring intensity distribution in Fig. 4c. The other way is to use an ultrahigh-power laser to irradiate on the target material, which further induces the photoelectric effect and makes electrons escape from the rear-side surface of the target material. Ref. [69] shows that the terahertz radiation with a field strength up to 100 GV/m in Fig. 4d is produced by two-color, ultrashort optical pulses interacting with under-dense helium gases at ultrahigh intensities.

**Beam diagnostics based on transition radiation**

In other to facilitate the implementation of high-gain free-electron laser or high-power electron beam, the shape of electron beams should be monitored. The coherent transition radiation offers a feasible method to diagnose the electron beam in high-energy equipment. This method mainly relies on the measurement of angular spectral energy density [169]. To be specific, the angular spectral energy density $\frac{d^2W_{\text{total}}}{d\omega d\theta}$ of the coherent transition radiation [169] is

$$\frac{d^2W_{\text{total}}}{d\omega d\theta} \cong N^2 F(\omega)\frac{d^2W_{\text{single}}}{d\omega d\theta} \qquad (2)$$

where $\frac{d^2W_{\text{single}}}{d\omega d\theta}$ is the angular spectral energy density for the transition radiation from a single electron, and $F(\omega)$ is a function of longitudinal and transverse distribution parameters for the electron beam after the Fourier transformation. With the knowledge of $\frac{d^2W_{\text{single}}}{d\omega d\theta}$ and $\frac{d^2W_{\text{total}}}{d\omega d\theta}$, the information of electron beams, namely $F(\omega)$, can be straightforwardly obtained in the experiment.

The beam diagnosis based on transition radiation was proposed in 1975 by Ref. [171], which uses two parallel foils. The phase and energy information of electron bunches can be inferred from the angular distribution of the interference pattern of transition radiation. In the following decades, more optical components (e.g. wire grid, Michelson interferometer) are adopted to provide not only the longitudinal and transverse distributions of electron beams but also their divergent angle with the help of coherent transition radiation [72-76].

Figure 5 shows a variety of setups for beam diagnosis, which are widely used in high-energy experiments and facilities, such as free-electron laser, wake-field acceleration, and large particle collider. When the electron beam of 42 MeV passes through an aluminum foil, the coherent transition radiation with a millimeter or submillimeter wavelength is observed in Fig. 5a [72]. The length of electron beams is measured by using a 45°-tilted foil and a polarizing Michelson interferometer in Fig. 5b, because the spectral power of light emission is dependent on the degree of coherency, which strongly relates to the beam size [73]. More precise measurement for the electron beam is proposed in AWAKE experiments through the usage of seeded self-modulation as shown in Fig. 5c [74]. At the CLARA facility, the coherent transition radiation is used to diagnose the longitudinal beam profile for the dielectric wakefield accelerator and coherent Cherenkov diffraction radiation [75]; see the setup in Fig. 5d.

**Excitation of surface wave by using transition radiation**

While surface waves can mold the flow of light at the subwavelength scale, their excitation is generally difficult, due to their momentum mismatch with propagating waves. In addition to the conventional schemes like gratings or prism matching, the transition radiation provides a powerful scheme to excite surface waves [77-84], especially for those with extremely-high spatial confinement such as graphene plasmons [78,172-175].

In 1957, Ritchie theoretically proposed a mechanism to excite surface waves by using transition radiation [176]. This prediction was observed in experiments in 2006 [77] by using an electron beam of 50 keV injected onto a flat gold surface as shown in Fig. 6a. In 2017, Ref. [78] theoretically revealed a splashing transient of graphene plasmons launched by swift electrons. During the process of transition radiation, a jet-like rise of excessive charge concentration emerges and is analogous to the hydrodynamic Rayleigh jet in a splashing phenomenon before the launching of ripples. When the free electrons interact with a mesoscopic structure composed of an array of nanoscale holes in a gold film, the surface waves are firstly excited and soon transformed into propagating waves [79]. As a result, ultrashort chirped electromagnetic wave packets could be created under the irradiation of 30-200 keV electron beams [79]. In 2019, the interaction between free electrons and photonic topological crystals is experimentally studied in Ref. [80]. The robust edge states are excited and observed, when a charged particle passes from one photonic crystal into another one.

**Outlook**

Despite the numerous applications enabled by the transition radiation, the transition radiation itself still suffers from low intensity and low directionality, especially if the charged particles have ultra-low energy. How to strengthen the particle-interface interaction and control the transition radiation in the desired way remains a long-standing challenge that is highly sought after. While metamaterials, 2D materials, and photonic crystals have been applied to tackle this challenge, there is still plenty of room to tailor the transition radiation by exploiting artificially-engineered materials and nanostructures. For example, we show in Fig. 7 that the gain materials, tilted hyperbolic materials, meta-boundaries, spatial-temporal materials, and layered structures with a periodic or random stacking might offer an enticing platform to enhance the particle-interface interaction and to achieve exotic features of transition radiation.

The gain material in Fig. 7a in practice can be implemented, for example, by using negative-resistance components [177-180] (e.g. microwave tunnel diodes) and optically pumped dye molecules [181,182] (e.g. Rhodamine 800 dye molecules). While gain materials could provide a universal way to amplify the light emission, the free-electron radiation from gain systems, including the transition radiation, has been rarely discussed. Particularly, the influence of optical gain and the slab thickness on the directionality of transition radiation remains elusive, while a larger optical gain is generally thought to have a larger enhancement of the intensity of transition radiation. More rich physics of free-electron radiation could be expected in systems simultaneously with optical gain and optical loss, such as those with parity-time symmetry [183-187]. However, the influence of the interplay between optical gain and optical loss on the transition radiation has never been explored before.

The hyperbolic material [188-196] is a uniaxial material that has a hyperbolic iso-frequency contour and is featured with an extraordinarily-large photonic density state. Due to these unique features, hyperbolic materials can greatly enhance both the particle-matter interaction and the particle-interface interaction. However, once these high-$k$ hyperbolic modes are excited, they cannot be safely coupled into free space, partly due to the existence of total internal reflection and material losses. How to overcome this issue becomes crucial for the development of novel on-chip light sources based on ultralow-energy electrons. To mitigate this issue, the photonic hyper-crystals are experimentally studied in Ref. [197]. The photonic hyper-crystals have combined virtues of strong light outcoupling in photonic crystals and large photonic density-of-states in hyperbolic metamaterials. Moreover, the broadband enhancement of on-chip photon extraction is achievable by using tilted hyperbolic metamaterials in Fig. 7b [198], since their eigenmodes now become momentum-matched with propagating waves in free space. Due to the exotic features of photonic hyper-crystals [197,199,200] and tilted hyperbolic metamaterials [198], the transition radiation from these exotic materials deserves more in-depth exploration.

The judicious design of electromagnetic boundary could provide a key route to control the free-electron radiation, including the transition radiation. Due to the recent advent of two-dimensional materials (e.g. graphene [201-208], $h$BN [209-217], twisted photonic structures [218-227]) and metasurfaces [89, 228-235], the boundary is uniquely featured with a surface conductivity, which can be rather complex but provide an extra degree of freedom to regulate the free-electron radiation. Without loss of generality, the boundary with a non-zero surface conductivity is termed as the meta-boundary in Fig. 7c [236]. According to the electromagnetic boundary conditions, the meta-boundary could be categorized into four types, including isotropic, anisotropic, biisotropic and bianisotropic meta-boundaries [236]. While the transition radiation from the simple isotropic meta-boundary (e.g. graphene) has been extensively studied [78,84,140-142], the transition radiation from more complex meta-boundaries remains largely unexplored and awaits more systematic investigation. Due to the powerfulness of metamaterial and meta-boundaries, the manipulation of free-electron radiation (e.g. transition radiation) at will might be enabled by combining meta-boundaries and metamaterials.

Spatial-temporal materials in Fig. 7d, such as photonic-time crystal and temporal metamaterials, have their optical response dependent both on space and time [237-239]. The emergence of spatial-temporal materials gets rid of fundamental limitations presented in space-engineered media and enable many counterintuitive phenomena, such as magnet-free nonreciprocity [240] and Cherenkov radiation in the vacuum [241,242]. Therefore, the spatial-temporal material in principle can provide a versatile platform to tailor the transition radiation. For example, Ref. [243] showed that a swift electron moving in a photonic-time crystal spontaneously emits radiation. When associated with momentum-gap modes, the process of free-electron radiation is exponentially amplified by the modulation of the refractive index [243]. Actually, the exploration of free-electron radiation from spatial-temporal material is still in its infancy.

Due to the appearance of multiple interfaces in layered structures, the transition radiation from each interface of the layered structure may interfere constructively or destructively. This way, through the judicious structural design, the layered structure may provide an extra degree of freedom to tailor the particle-interface interaction. For example, if the layered structure has a periodic stacking in Fig. 7e, the constructive interface of transition radiation from each interface can be regarded as the excitation of high-$k$ Bloch modes inside the photonic crystals. As advantageous to the high-$k$ modes in hyperbolic materials, the excited high-$k$ Block modes in layered structures with a periodic stacking might be safely coupled into free space, due to the umklapp scattering. Therefore, the particle-interface interaction is promising to efficiently extract the information from ultralow-energy electron, which however remains further studies. Moreover, the dispersion is unavoidable in layered structures with a periodic stacking, either from the material-induced dispersion or the structural-periodicity-induced dispersion. How to reduce or even eliminate the dispersion in layered structures so that the dispersionless resonance transition radiation can be created is still a challenge in science and technology.

If the layered structure has a non-periodic stacking in Fig. 7f, the interference of transition radiation from each interface becomes more random but offers more degrees of freedom to tailor the particle-interface interaction. Due to the disappearance of eigenmodes or Bloch modes in the randomly-stacked layered structures, we can no longer predict the behavior of transition radiation simply from the optical response of the randomly-stacked layered structures. To enable the theoretical prediction, the rigorously analytical solution of transition radiation from the layered structures, although being quite tedious, is mandatory. As a result, whether we can achieve the constructive interference of transition radiation from each interface at a specific radiation angle remains elusive. From the perspective of applications, the associated resonance conditions for transition radiation from the randomly-stacked layered structures are highly wanted. Moreover, due to the non-period nature of the randomly-stacked layered structures, whether they can be adopted to achieve broadband dispersionless resonance transition radiation is also worth further investigation.

## Competing interests

The authors declare no competing interests.

## Acknowledgement

X.L. acknowledges the support partly from the National Natural Science Fund for Excellent Young Scientists Fund Program (Overseas) of China, the National Natural Science Foundation of China (NSFC) under Grant No. 62175212, Zhejiang Provincial Natural Science Fund Key Project under Grant No. LZ23F050003, the Fundamental Research Funds for the Central Universities (2021FZZX001-19), and Zhejiang University Global Partnership Fund. H.C. acknowledges the support from the Key Research and Development Program of the Ministry of Science and Technology under Grants No. 2022YFA1404704, 2022YFA1404902, and 2022YFA1405200, the National Natural Science Foundation of China (NNSFC) under Grants No.11961141010 and No. 61975176. J.C. acknowledges the support from the Chinese Scholarship Council (CSC No. 202206320287).

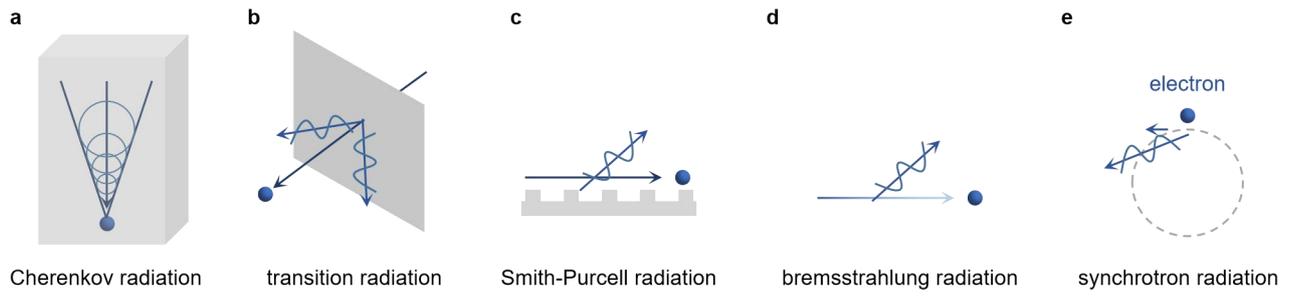

**Fig. 1 Schematic of different types of free-electron radiation. a**, Cherenkov radiation. **b**, Transition radiation. **c**, Smith-Purcell radiation. **d**, Bremsstrahlung radiation. **e**, Synchrotron radiation.

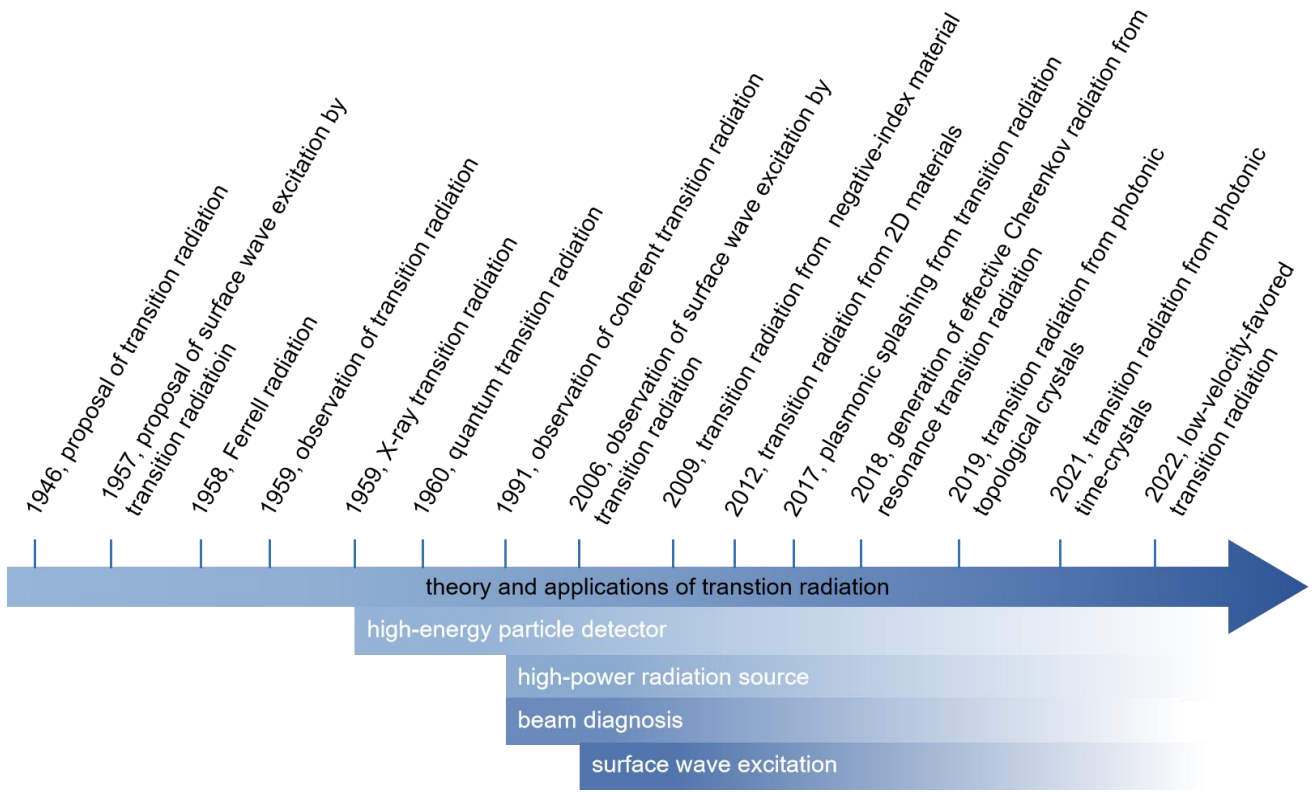

**Fig. 2 Brief history of transition radiation, along with its typical applications.**

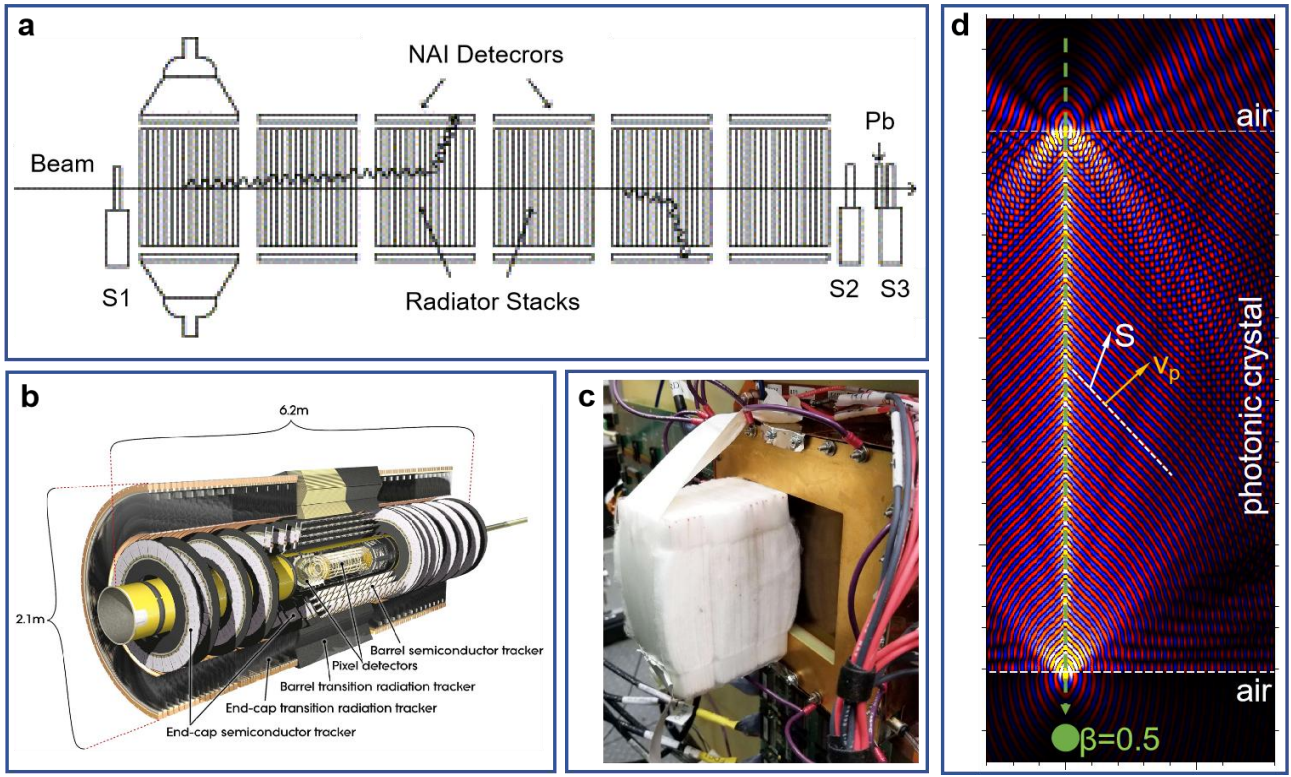

**Fig. 3 High-energy particle detectors based on transition radiation. a**, An inorganic scintillator-based transition radiation detector at the super proton synchrotron (SPS) of CERN [60]. **b**, Cutaway view of the ATLAS inner barrel detectors [61]. **d**, Transition radiation detector based on the gas electron multipliers (GEM) technology [64], which can improve the detector performance to identify multiple particles. **d**, Design of Cherenkov detectors by using the resonance transition radiation [150].

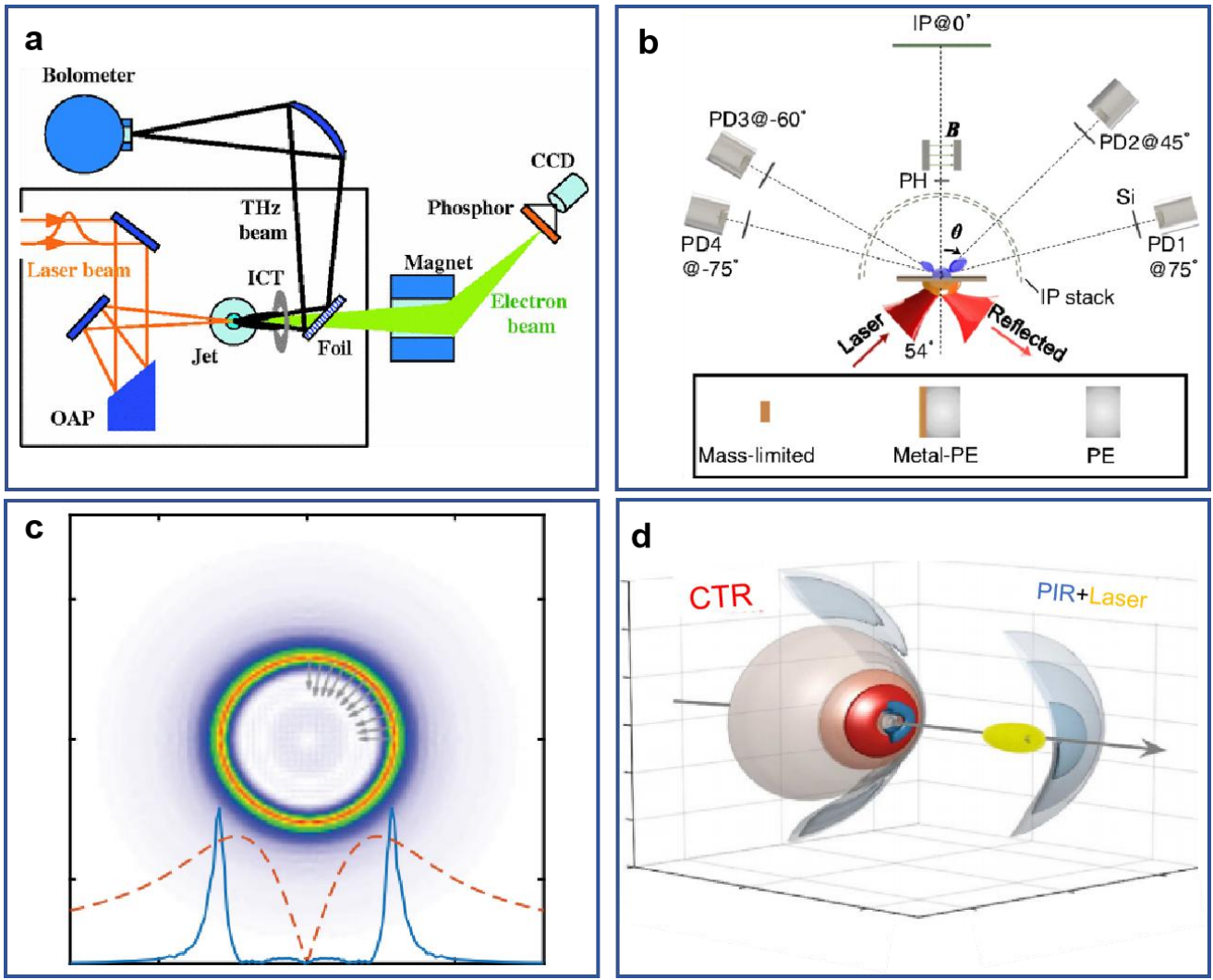

**Fig. 4 High-power radiation source by exploiting the coherent transition radiation (CTR). a**, High power terahertz emission from a laser-plasma accelerated electron bunch [66]. **b**, Demonstration of terahertz emissions with energy of sub-mJ/pulse [67], when the laser-produced electron beam passes through the rear dielectric-vacuum interface. **c**, Coherent transition radiation by the wakefield-accelerated electrons to yield a field strength of ~100 GV/m [68]. **d**, Terawatt attosecond pulses from ultraviolet transition radiation [69].

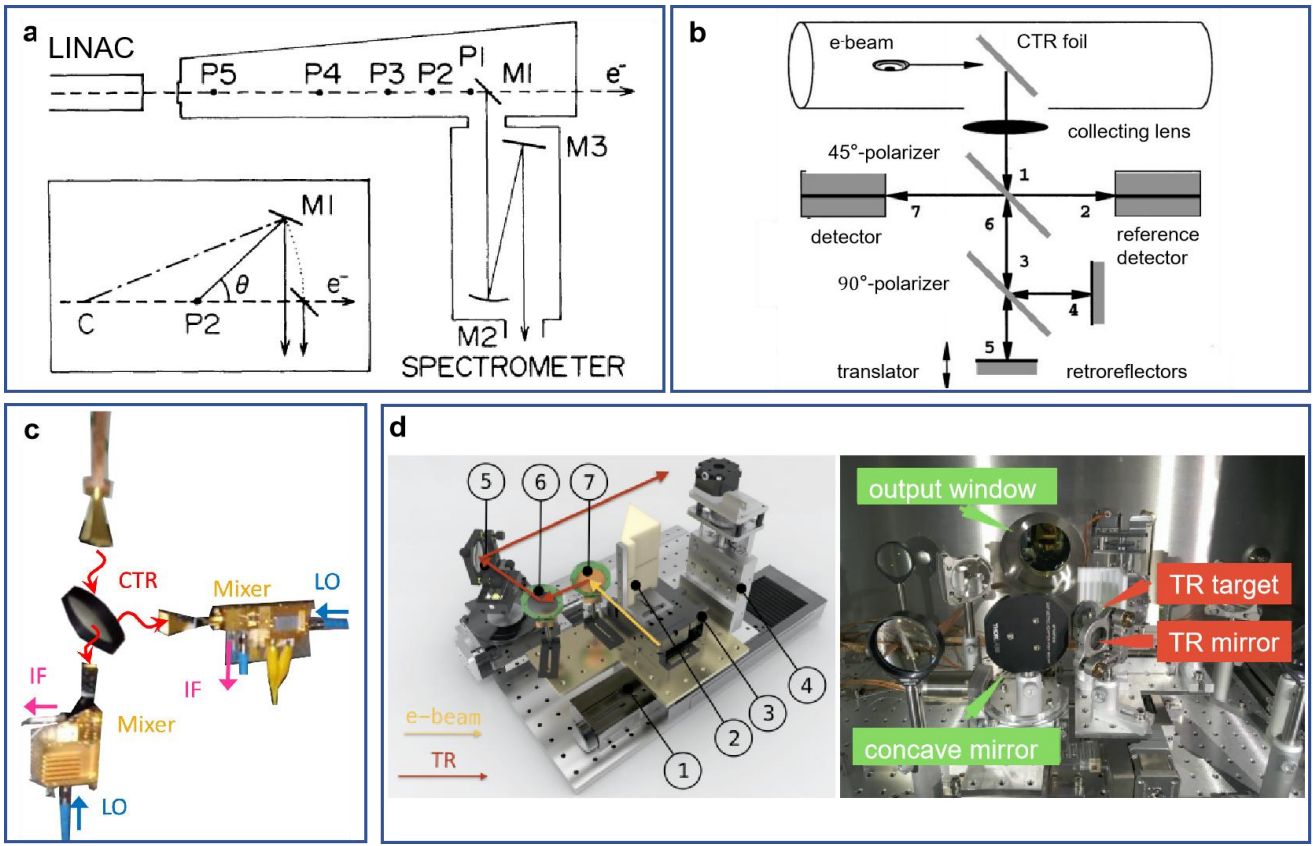

**Fig. 5 Beam diagnosis technology based on transition radiation. a**, Schematic of the experimental setup to observe the coherent transition radiation at millimeter and submillimeter wavelengths [72]. **b**, Bunch length measurement of picosecond electron beams [73]. **c**, A waveguide-integrated heterodyne diagnostic at the AWAKE (Advanced WAKEfield Experiment) of CERN via the coherent transition radiation [74]. **d**, A longitudinal beam profile monitor based on the coherent transition radiation in CLARA (Compact Linear Accelerator for Research and Applications) [75].

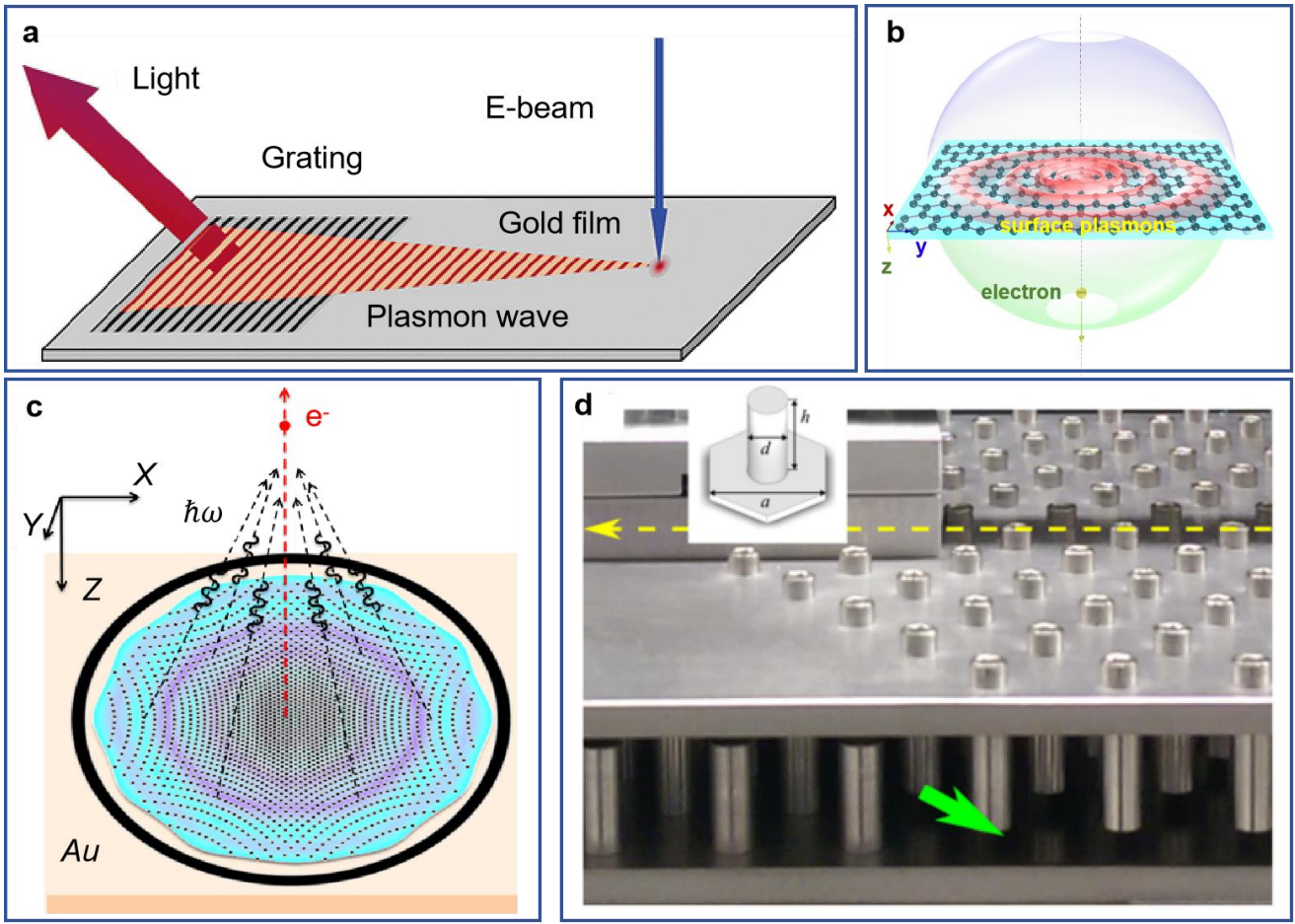

**Fig. 6 Excitation of surface waves by transition radiation. a**, A free-electron beam is injected into a gold surface to create a highly-localized source of surface plasmons [77], which would be further scattered into free space after their interaction with the grating. **b**, Plasmonic splashing from transition radiation [78]. **c**, Femtosecond photon bunches from the coherent scattering of hyperbolic surface waves [79], which are excited by the transition radiation. **d**, Generation of edge waves when an electron beam passes through an interface between two different photonic crystals [80].

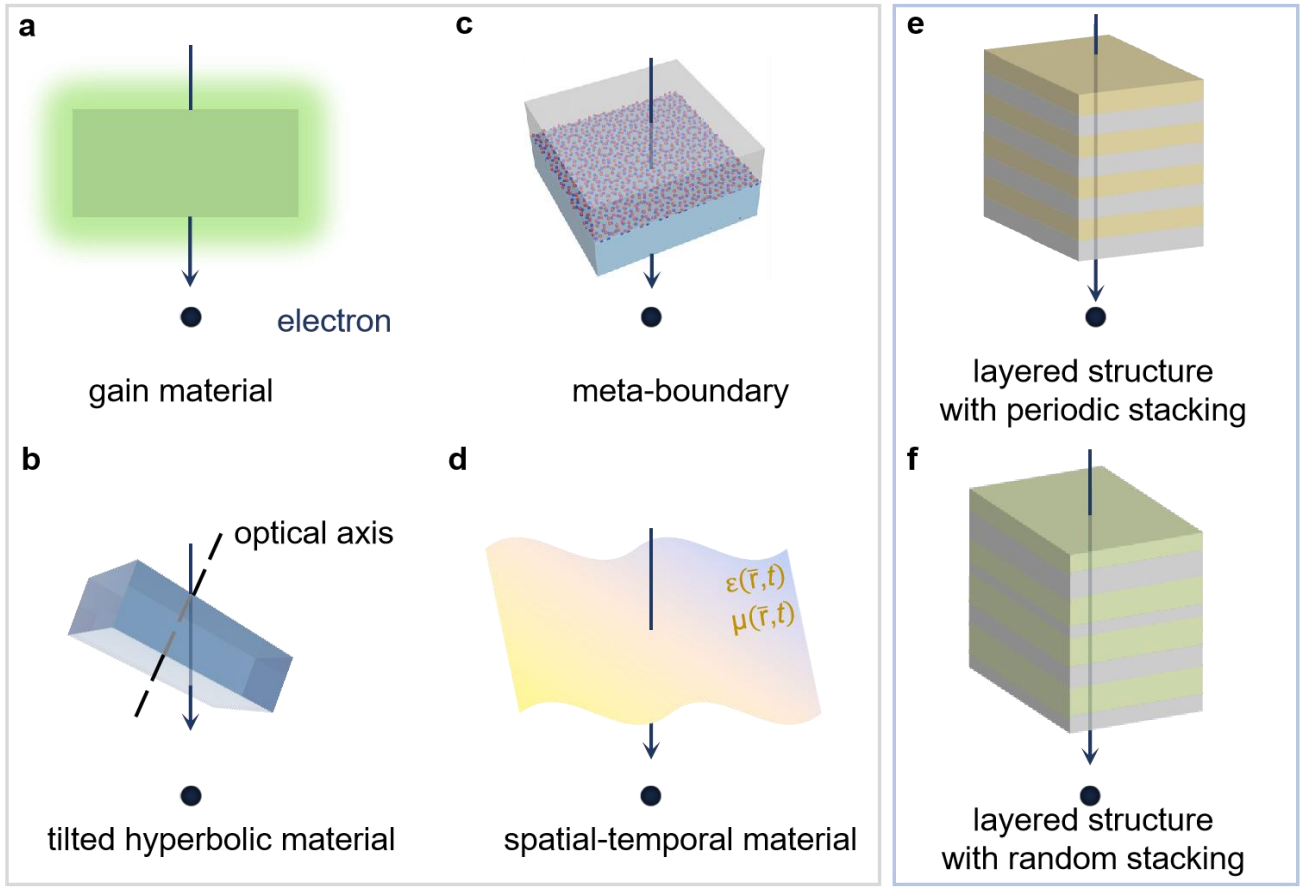

**Fig. 7 Manipulation of transition radiation via artificially-engineered materials or nanostructures. a**, Gain material. **b**, Titled hyperbolic material. **c**, Meta-boundary, such as twisted bilayer graphene. **d**, Spatial-temporal material. **e**, Layered structure with a periodic stacking. **F**, Layered structure with a non-periodic (or random) stacking.